# Facilities for the Energy Frontier of Nuclear Physics

John M. Jowett

**Abstract**

The Relativistic Heavy Ion Collider at BNL has been exploring the energy frontier of nuclear physics since 2001. Its performance, flexibility and continued innovative upgrading can sustain its physics output for years to come. Now, the Large Hadron Collider at CERN is about to extend the frontier energy of laboratory nuclear collisions by more than an order of magnitude. In the coming years, its physics reach will evolve towards still higher energy, luminosity and varying collision species, within performance bounds set by accelerator technology and by nuclear physics itself. Complementary high-energy facilities will include fixed-target collisions at the CERN SPS, the FAIR complex at GSI and possible electron-ion colliders based on CEBAF at JLAB, RHIC at BNL or the LHC at CERN.

# Facilities for the Energy Frontier of Nuclear Physics

John M. Jowett

European Organization for Nuclear Research (CERN), CH-1211 Geneva 23

John.Jowett@cern.ch

**Abstract**. The Relativistic Heavy Ion Collider at BNL has been exploring the energy frontier of nuclear physics since 2001. Its performance, flexibility and continued innovative upgrading can sustain its physics output for years to come. Now, the Large Hadron Collider at CERN is about to extend the frontier energy of laboratory nuclear collisions by more than an order of magnitude. In the coming years, its physics reach will evolve towards still higher energy, luminosity and varying collision species, within performance bounds set by accelerator technology and by nuclear physics itself. Complementary high-energy facilities will include fixed-target collisions at the CERN SPS, the FAIR complex at GSI and possible electron-ion colliders based on CEBAF at JLAB, RHIC at BNL or the LHC at CERN.

## 1. Introduction

The turn of this century coincided with a shift in the accelerator technology used to explore the energy frontier of nuclear physics.

Since Rutherford's α-Au experiment, nine decades previously, laboratory experiments were conducted with beams of nuclei accelerated at other, stationary, nuclei in targets. Meanwhile, elementary particle physics, as it gradually demarcated itself from nuclear physics in the 1960s, extended its energy frontier significantly by adopting the technique of colliding particle beams. Yet, apart from a very brief interlude in the late 1970s when deuterons (D or $^{2}H^{1+}$)) and α-particles ($^{4}He^{2+}$) were collided at the CERN Intersecting Storage Rings (ISR) [1,2], only electrons, protons and their anti-particles, in various combinations, were ever collided in storage rings.

This changed in 2000 with the advent of the first dedicated nucleus-nucleus (A-A) collider, the Relativistic Heavy Ion Collider (RHIC) at Brookhaven National Laboratory (BNL). From that time on, "high-energy physics" could no longer be considered synonymous with "elementary particle physics". Numerous papers in this conference describe the physics of nuclear matter at extreme temperature and energy density, the Quark-Gluon Plasma, that continues to be explored by RHIC and its detectors. All the matter identified in our present universe is understood to have existed in this form in the first microseconds after the Big Bang.

A second heavy-ion collider, the Large Hadron Collider (LHC) at CERN is already colliding protons and will start to collide lead nuclei later this year. With its much higher energy and exceptionally capable detectors, it will continue to mine this rich seam of fundamental physics. Again, the expectations are abundantly documented in contributions to this conference.

While this paper will focus on the accelerator physics, performance and evolution of these two colliders, it should not be forgotten that new opportunities to study heavy-ion physics with fixed targets and/or colliding beams at lower energies are also being prepared at NICA at JINR, FAIR at GSI and CERN's SPS.

To date, only one lepton-hadron collider, HERA at DESY, has been built and it focused exclusively on e±-p collisions. The untapped physics opportunities afforded by electron-ion (e-A) colliders have led to a surprising variety of proposals, most conceived as building on existing accelerator facilities. I shall conclude with a brief survey of present ideas for high-energy e-A colliders.

1.1. Generalities

To fix terminology and notation, let us review some essential facts about the performance of a hadron collider; further explanation and details can be found in many references, eg, the handbook [3]. The beam energy (is traditionally quoted as $E_N$ GeV per nucleon where the energy of a fully-stripped relativistic ion is

$$E = AE_N \approx ZE_p \tag{1.1}$$

and $E_p$ is the energy that a proton would have in the same magnetic ring; thus we can quote the energy of a gold nucleus in RHIC as $E = 100 A$ GeV $= 250 Z$ GeV $= 19.7$ TeV. To simplify, we consider only rings with bunched beams, the $k_b$ bunches being supposed of equal intensity, $N_b$, the number of particles (ions, nuclei) per bunch. In most hadron colliders, the beams are meant to be round, in that not only are the horizontal and vertical geometric emittances equal:

$$\varepsilon_x = \varepsilon_y = \varepsilon = \frac{\varepsilon_n}{\sqrt{\gamma^2 - 1}}, \tag{1.2}$$

but the betatron amplitude functions are also equal by design at the interaction point (IP, starred symbols),

$$\beta_x^* = \beta_y^* = \beta^*. \tag{1.3}$$

Then the vertical and horizontal RMS beam sizes are equal at the IP:

$$\sigma_x^* = \sqrt{\beta_x^* \varepsilon_x} = \sigma_y^* = \sqrt{\beta_y^* \varepsilon_y} = \sigma^* \simeq \sqrt{\frac{\beta^* \varepsilon_n}{\gamma}} \quad (\gamma \gg 1). \tag{1.4}$$

The so-called normalised emittance $\varepsilon_n$ is most frequently quoted in parameter lists, in units of $\mu$m and, ideally, remains constant as the energy of the collider is ramped up. Note that we are following the convention of RMS emittances, as at the LHC; at RHIC, for example, the so-called 95% emittance is usually quoted and is a factor 6 larger [3]. These formulas are strictly correct only when the bunches have Gaussian charge distributions.

If all bunches in both beams are identical in $\varepsilon_n$ and population, $N_b$, the event rate for a process with cross-section $\sigma$ is $L\sigma$, where the luminosity is given by

$$L = \frac{N_b^2 k_b f_0}{4\pi \sigma_x \sigma_y} F = \frac{N_b^2 k_b f_0 \gamma}{4\pi \varepsilon_n \beta^*} F(\theta_c) \tag{1.5}$$

where $f_0$ is the revolution frequency, and the factor

$$F = 1 / \sqrt{1 + \left(\frac{\theta_c \sigma_z}{2\sigma^*}\right)^2} = \left[1 + \frac{\gamma (\theta_c \sigma_z)^2}{4\beta^* \varepsilon_n}\right]^{-1/2} \tag{1.6}$$

takes account of the geometric reduction in luminosity when bunches of RMS length $\sigma_z$ collide with a half-crossing angle $\theta_c$. Depending on the bunch filling pattern, the number of collisions per turn may be less than $k_b$ at a given IP. Some generalisations of these formulas are fairly obvious; others are assembled conveniently in [3].

Thus, to achieve high collider performance, we must strive to create intense beams of small emittance in the ion source and/or cooling rings, then preserve the intensity and emittance through the chain of injectors, injection into the collider, the ramp to full energy and finally the "squeeze", the process in which the beam optics around the collision point is modified to achieve the smallest possible $\beta^*$. There are many challenges on the way, those scaling with high powers of the nuclear charge being particularly troublesome for heavy-ion colliders.

Luminosity itself removes particles from the beam, the so-called "burn-off", with contributions from each IP: $\dot{N}_b = -\sum_{IP} L\sigma_T + N_b/\tau_s$, where $\sigma_T$ is the total cross section for all processes removing particles from the beam and $\tau_s$ is the beam lifetime due to other single-beam processes. The luminosity (1.5) will decay with an instantaneous time-constant

$$\frac{1}{\tau_L} = -\frac{\dot{L}}{L} = 2\frac{\sum_{IP} L\sigma_T}{N_b} + \frac{2}{\tau_s} + \frac{\dot{\varepsilon}_n}{\varepsilon_n} + \frac{\dot{\beta}^*}{\beta^*} \tag{1.7}$$

(Note that $L, \tau_s, \dot{\varepsilon}_n \ldots$ are more or less complicated functions of $N_b, \varepsilon_n, \sigma_z$ and other beam parameters so (1.7) actually represents a system of coupled ordinary differential equations [4,5].) The time-integrated luminosity performance of a collider can be discussed in terms of the factors in (1.5) and the terms in (1.7).

Unfortunately, the hadronic component of $\sigma_T$, the main physics interest, is dwarfed by terms depending on very high powers of *Z*, due to ultraperipheral electromagnetic interactions which cause high initial luminosity to decay rapidly [4,5,6]. Since the time taken to dump the beams, recycle, refill and bring the collider back to physics conditions is irreducible beyond a certain point, the only way to gain in average luminosity is to make $\dot{\varepsilon}_n < 0$ (beam cooling).

The last term in (1.7) usually vanishes; however if there is some limit on the instantaneous luminosity, but not the intensity, then it also pays to increase the initial $N_b$ and adjust the optics to make $\beta^* \propto \sqrt{N_b}$, maintaining the relationship as time goes on until the lower limit on $\beta^*$ is reached. Although this is one of the original proposals [7] for luminosity-levelling, it has not really been tried so far; others, such as varying an offset between the beams, require a more general starting formula than (1.5).

While p-p or p-$\bar{p}$ colliders advance through an essentially two-dimensional performance parameter space (energy and luminosity), ion colliders are required also to provide collisions between varying species of nuclear charge $Z_1, Z_2$ and baryon numbers $A_1, A_2$ (in this they are less flexible than their stationary target predecessors). Comparisons of centre-of-mass energy, central rapidity and luminosity must then be made on the basis of the nucleon-nucleon values,

$$\sqrt{s_{NN}} \approx 2c E_p \sqrt{\frac{Z_1 Z_2}{A_1 A_2}}, \quad y_{NN} = \frac{1}{2}\log\frac{Z_1 A_2}{A_1 Z_2}, \quad L_{NN} = A_1 A_2 L. \tag{1.8}$$

where $E_p$ is the beam energy of protons, assuming the same magnetic field for each beam (N.B. this is not the case for RHIC's D-Au configuration). Thus, the design Pb-Pb luminosity, $L = 10^{27}$ cm$^{-2}$s$^{-1}$, of the LHC is comparable with a p-p luminosity of $L_{NN} = 4.3\times 10^{31}$ cm$^{-2}$s$^{-1}$, especially as concerns, say, the production of high-transverse momentum hadronic jets.

## 2. The Relativistic Heavy Ion Collider (RHIC)

For the moment, RHIC is the only ion collider in operation. Remarkably, it is also the only high energy polarized proton collider ever to operate and is likely to hold that title indefinitely. However I shall not discuss that aspect of its operation in this paper; the latest update can be found in [8]. Another major aspect of the RHIC programme, not discussed here, is the low-energy nucleus-nucleus collisions used in the search for the QCD critical point (a difficult, low-energy frontier for accelerator physics).

The collider is composed of two, largely separate, 3.8 km circumference rings of superconducting magnets and presently houses two large collider experiments, PHENIX and STAR. Smaller experiments, BRAHMS and PHOBOS have also operated in the past.

Heavy-ion collisions at RHIC are mostly gold-gold ($^{197}$Au$^{79+}$, the only stable isotope) but D-Au and copper-copper ($^{63}$Cu$^{29+}$) collisions have also been provided. Figure 1 shows the steady increase in performance over the past decade, reflecting the accumulation of operational experience and a series of judicious and effective measures to improve performance.

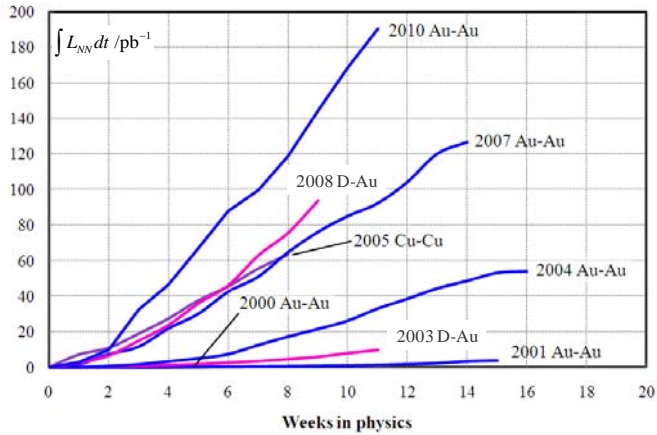

Figure 1: RHIC nucleon-nucleon integrated luminosity, as defined in Eq. (1.8), delivered to the PHENIX experiment in each of the runs that have taken place so far, from [9]. All beam species are shown, save protons, and time is measured in weeks from the start of physics data-taking in each run (this excludes set-up time).
The average Au-Au luminosity in a store is now $\langle L \rangle = 2.\times 10^{27}$ cm$^{-2}$s$^{-1}$, an order of magnitude beyond the original RHIC design luminosity [10].

2.1. Limitations on the performance of RHIC

In beam physics, the term *intra-beam scattering (IBS)* refers to multiple small-angle Coulomb scattering processes that occur between particles oscillating inside the beam [3]. The consequent emittance growth rates are roughly inversely proportional to the emittances themselves. This causes a diffusion of particles in the transverse and longitudinal phase space resulting initially in a growth of the emittances and bunch length (3$^{rd}$ term in (1.7)). After a while, particles in the beam tails may start to reach the collimators delimiting the transverse phase space or the boundary of stable longitudinal motion ("RF bucket"). They will then start to be lost from the beam or circulate outside the buckets (2$^{nd}$ term in (1.7)). These growth times and loss rates contain the factor $Z^4/A^2$ from the Coulomb cross-section so IBS is important for high-Z nuclear beams. Together with space charge, it is one of the reasons why ions are kept in lower charge states in injector synchrotrons where maximum energy is not yet the priority.

In the early years of RHIC operation, IBS was the dominant luminosity limit, blowing up the transverse emittances and leading to the rapid decay of luminosity exemplified in Figure 2; the growth of the initial small emittances is particularly rapid and the ratio of average to peak luminosity is low.

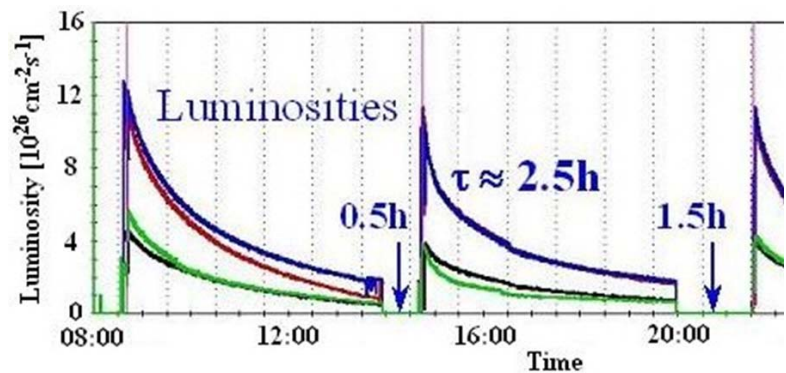

Figure 2: Luminosities of the four RHIC experiments during part of a day in 2004 (before any implementation of beam cooling). The rapid luminosity decay is mainly due to the emittance increase from IBS. Eventually the beams are dumped and the collider refilled, a process that takes about 30 or 90 minutes in the cases shown; see also [5].

As equation (1.5) shows, reduction of $\beta^*$ is also a path to higher luminosity, *provided* the bunch length $\sigma_z$ can be kept short enough (equation (1.6)). The present limit $\beta^* = 0.75$ m comes from the difficulty of correcting the non-linear variation of the betatron frequencies across the momentum spread in the bunches [10,9]. This may improve in future with lattice designs incorporating the necessary chromatic corrections.

During acceleration, the Au beams in RHIC have to cross the transition energy, where the first-order variation of revolution frequency of individual particles with momentum vanishes and they are particularly sensitive to intensity-related instabilities. Some of these are driven by the impedance presented by the vacuum envelope of the beam while others are related to the formation of electron clouds by secondary emission from the chamber surfaces. Such effects lead to intensity losses, predominantly of the later bunches in the trains. Efforts continue to combat these effects with feedback systems and the reduction of the secondary emission yield of the arc chambers (the straight sections are already NEG-coated).

2.2. Upgrades improving performance

Although the technique of stochastic cooling has long been established for coasting (ie, unbunched) beams, it is only relatively recently that it has been successfully applied to high energy bunched beams. Longitudinal stochastic cooling was implemented in RHIC [11] in 2007. More recently (see Figure 3) first successes with cooling of the vertical emittance have been reported and it is already very effective in countering the effects of IBS [10].

Bunched beam stochastic cooling will soon be implemented in all three planes for both beams and is expected to provide a factor 4 increase (already partly realised) in average luminosity. In addition, modifications to the RHIC RF system (stronger longitudinal focusing from new 56 MHz RF cavities, suppression of common cavities) will reduce longitudinal losses from IBS diffusion. At later stages of a fill, the luminosity decay due to intensity burn-off in collisions will be largely compensated by the reduced emittances achieved by these measures.

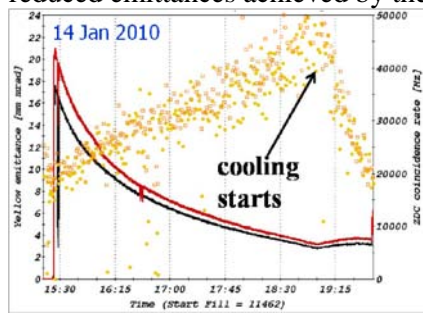

**Figure 3:** The first vertical stochastic cooling in RHIC, earlier this year. The yellow points are emittance measurements from an ionization profile monitor (left scale) and show the initial IBS growth being reversed when the cooling is switched on. The luminosity signals measured by the Zero-Degree Calorimeters of the two experiments (red and black, right scale) show the recovery of luminosity. Further information and references can be found in [9], the source of this figure.

Our BNL colleagues are also commissioning a new Electron Beam Ion Source (EBIS) and pre-injectors which will eliminate the dependence on the ageing Tandem Van de Graaff's and provide new flexibility to vary species, including U beams. End-on collisions of these non-spherical nuclei will extend the high energy tail of energy densities produced in collision.

A recent, more detailed account of these and other upgrade plans is provided in [9] and Table 1 provides a comparison of RHIC and LHC parameters.

**3. The Large Hadron Collider (LHC)**

The general design of the LHC and its heavy-ion injector chain can be found in [12]. The ECR ion source, Linac 3 and the accumulation and electron-cooling ring LEIR were built especially for heavy ions. Together with the PS and SPS synchrotrons they constitute the heavy-ion injector chain which has been commissioned in the past few years and is close to achieving its design performance [14].

Three highly capable and complementary detectors, ALICE, ATLAS and CMS will study the heavy-ion collisions [20], as described in other papers at this conference.

Table 1 shows that, even at the initial half-nominal energy, the LHC will extend the energy frontier for laboratory nuclear collisions a factor 13.7 (later up to 28); this seems to be the largest energy step ever made by any collider over its predecessor.

The heavy-ion runs, set-up time and physics operation together, will typically last for 1 month at the end of each year. There will therefore be a high premium on operational efficiency: refer back to Figure 1, which does not include set-up time, and imagine each run truncated at ~3.5 weeks. One might reasonably ask whether heavy-ion physics will ever really get started at the LHC?

Nevertheless, the first Pb-Pb run is planned for 2 November and we project that it should yield an integrated luminosity of at least $3 \,\mu b^{-1}$. The LHC has started this year at reduced energy and moderate values of $\beta^*$ in a staged approach to commissioning and these will be taken over for the heavy-ion run (see Table 1); the energy will be increased after the first long shutdown [15].

**Table 1:** Comparison of principal parameters of RHIC and LHC. The two columns for RHIC are those achieved recently this year and those projected after completion of the present upgrades in 2012 [9]. For LHC we give presently estimated performance in the first run later this year (although we are presently considering doubling the number of bunches) and the long-standing design parameters from 12]. Note that collimation efficiency and/or BFPP effects may limit the peak LHC luminosity to a lower value until dispersion suppressor collimators are installed.

| | | RHIC Au-Au | | LHC Pb-Pb | |
|---|---|---|---|---|---|
| Quantity | Unit | Achieved | Upgrade | 2010 | Design |
| Ion Energy | TeV | $19.7 = 0.1A = 0.25Z$ | | $287 = 1.38A = 3.5Z$ | $574 = 2.76A = 7.Z$ |
| $\sqrt{s_{NN}}$ | TeV | 0.2 | | 2.76 | 5.51 |
| No. of bunches $k_b$ | | 111 | | 62 | 592 |
| Ions/bunch $N_b$ | | $1.1 \times 10^9$ | $1.0 \times 10^9$ | $7 \times 10^7$ | |
| Stored energy | MJ | 0.39 | 0.35 | 0.2 | 3.8 |
| Optics at IP $\beta^*$ | m | 0.75 | 0.5 | 3.5 | 0.5 |
| Emittance $\varepsilon_n$ | $\mu$m | 0.47 | 0.42 | 1.5 | |
| RMS bunch length | m | 0.3 | 0.3 | 0.08 | |
| $F$ (defined in (1.6) | | 0.93 | 0.88 | 1 | 0.98 |
| Peak luminosity $L$ | $10^{26}$ cm$^{-2}$s$^{-1}$ | 40 | 55 | 0.07 | 10 |
| Average lumi. | $10^{26}$ cm$^{-2}$s$^{-1}$ | 20 | 40 | ~0.02 | ~4 |

3.1. First injection in 2009 and plans for 2010

Commissioning of the LHC with proton beams started at a brisk pace in September 2008 but was interrupted after a few days by the accident which led to a year's shutdown for repairs [15]. The machine was reawakened with a first injection of lead nuclei from the heavy-ion injector chain. The initial uncorrected trajectory is shown in Figure 4. Some corrections were then applied but this is as far as we have gone with ion beams so far. Shortly afterwards we switched back to protons and observed that they reproduced, very precisely, the same initial trajectories as the Pb. This fact, that beams of the same magnetic rigidity, *p / Ze and injection conditions*, should follow exactly the same trajectories in the same *static magnetic* fields, is key to the strategy for rapidly switching back from protons to Pb ions that we intend to apply later this year. Of course, this does not apply to the effects of electric fields, the main practical consequence of which is that the RF frequency will have to be reduced by 5 kHz. Nevertheless, it means that the magnetic configuration of the machine, from injection, through the ramp and squeeze, will remain mostly valid, saving time after the species-switch. Some changes will have to be made, mainly to the ALICE collision configuration and some time will be spent to set up the collimation system and ensure proper machine protection. We believe we can establish conditions for Pb-Pb data taking in less than a week.

Meanwhile steady progress is being made with p-p collisions: today, the number of bunches per beam has been increased to $k_b = 10$ with intensity approaching $N_b = 10^{11}$ (already close to the design value and the luminosity is approaching $L = 10^{30}$ cm$^{-2}$s$^{-1}$. The first 100 nb$^{-1}$ of integrated luminosity has been delivered to the experiments in about 180 days of operation. Note that the parameters of the first heavy ion run (Table 1), 4 months from now, require us to go quickly to many more bunches than are injected today.

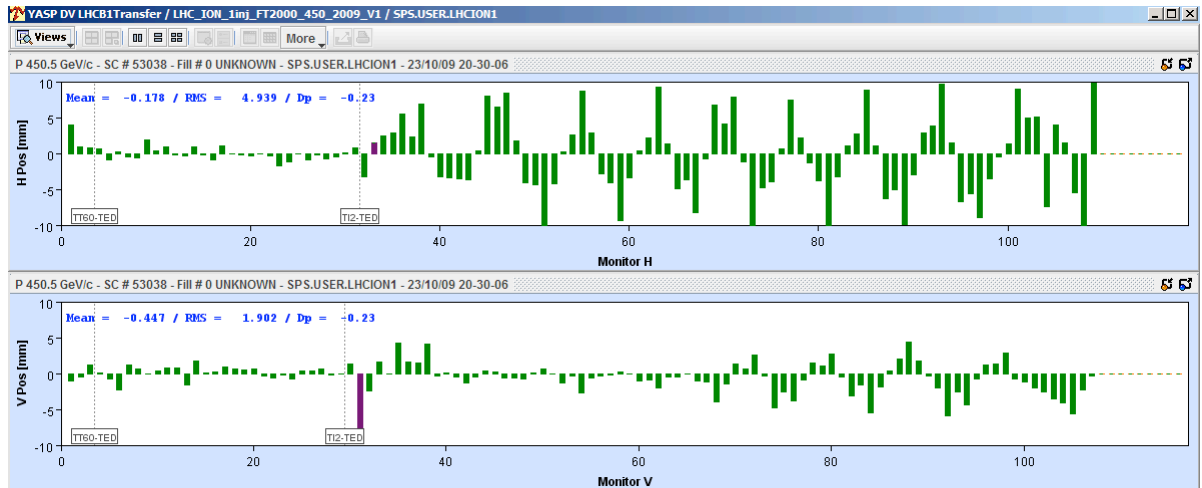

**Figure 4:** First Pb ions in the LHC: horizontal and vertical trajectories of the Pb beam injected into the LHC, in the evening of 23 October 2009, displayed as a function of beam-position monitor number. The first 30 monitors on the left are in the transfer line from the SPS and the remainder span two octants of the LHC, starting from the injection point close to the ALICE experiment.

3.2. Limitations on the performance of the LHC

Beyond the initial magnetic setup, there are very substantial differences between protons and heavy ions in the LHC. As at RHIC, IBS will be quite strong at injection and not negligible at the initial operating energy of 3.5 $Z$ TeV although we expect it to be countered by the emergence of synchrotron radiation damping at 7 $Z$ TeV [4,5]. This natural cooling effect varies as $E_p^3 Z^5 / A^4$, so is faster for lead nuclei than for protons.

Several years ago, it was realised [16] that the same ultraperipheral electromagnetic effects that are responsible for the short luminosity lifetime (cross sections of a few hundred barn) in heavy-ion collisions can also directly limit the maximum value of the luminosity. Processes such as bound-free pair production or two-neutron electromagnetic dissociation,

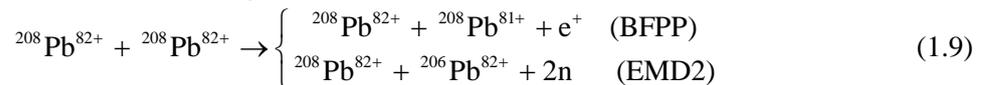

$$^{208}\text{Pb}^{82+} + {}^{208}\text{Pb}^{82+} \rightarrow \begin{cases} {}^{208}\text{Pb}^{82+} + {}^{208}\text{Pb}^{81+} + e^+ & \text{(BFPP)} \\ {}^{208}\text{Pb}^{82+} + {}^{206}\text{Pb}^{82+} + 2n & \text{(EMD2)} \end{cases} \quad (1.9)$$

manifest themselves as secondary beams of modified nuclei emerging from the IP and impinging on well-defined spots on the vacuum chamber [4]. At full LHC energy, the most complete calculations [17] suggest that the resulting heating can quench superconducting magnets at luminosities somewhat below the original design value. However there are many uncertainties related to the hadronic shower calculations, heat transfer in the liquid helium flowing through the magnet coils and their propensity to quench. The most promising counter-measure to increase the luminosity in an experiment would be the installation of so-called "cryo-collimators" at specific locations among the bending magnets closest to it. These require substantial modifications of the machine, including the moving of magnets to liberate space and can only be contemplated during long shutdowns. However one should also remember that, if the peak value is limited, the ratio of integrated to peak luminosity can be higher.

In p-p operation, the LHC relies on an elaborate two-stage collimation system, based on the diffractive scattering of protons in the primary collimator material, to clean the beam of halo particles. Hadronic fragmentation and electromagnetic dissociation of nuclei impinging on the primary collimators mean that the system works like a single-stage system and collimation efficiency is poorer, with a whole spectrum of nuclides being lost into superconducting magnets, with consequences—and possible counter-measures—similar to those mentioned above [4,12]. This constitutes a limit on total intensity rather than luminosity.

3.3. Hybrid collisions and other nuclei

Heavy-ion collisions at LHC will be mostly lead-lead ($^{208}$Pb$^{82+}$, the heaviest of all stable isotopes) but p-Pb and, later, lighter ion ($^{40}$Ar$^{18+}$) collisions are also foreseen. Argon and xenon beams will be made available in the next few years to serve the fixed target programme at the SPS; when their intensities are known, it will become possible to estimate the luminosity that might be attained with them in the LHC.

The hybrid p-Pb collisions are of crucial importance to the physics programme, both as a reference for interpretation of the nucleus-nucleus collisions and for their potential to illuminate the partonic structure of matter at low Bjorken $x$. They cannot be achieved in the same way that D-Au collisions were at RHIC [18]. There the revolution frequencies of the two beams could be kept equal by varying the magnetic fields of the two rings independently. This is precluded by the two-in-one magnet design of the LHC which imposes equal magnetic fields and unequal frequencies on the two beams at injection and during the ramp, until the energy is high enough that the differences can be absorbed by small, opposite displacements of the central orbits [19]. RHIC experience [18] suggests there may be difficulties with this although contrary arguments [19] suggest that an adequate luminosity of order $L \approx 1.5 \times 10^{29}$ cm$^{-2}$s$^{-1}$ is within reach; it must be recognized that some uncertainty about the feasibility of this collision mode remains.

**4. Electron-Ion Colliders**

There are three proposals for high-energy electron-ion colliders (as well as some at lower energy). So far none has been approved. In each case, the project is conceived as an extension of an existing facility which already provides one of the two beams. A recent survey [20] provides considerably more detail and references than we have space for here and includes a comprehensive table comparing the design parameters of these innovative and complex facilities.

4.1. eRHIC

Adding an electron beam to collide with the existing hadron beams is a very natural extension of the RHIC complex. Ideas for implementation have evolved and matured over a number of years and R&D on critical systems is well advanced, with a view towards demonstrating feasibility of the most critical systems. The present staged concept envisions
- MeRHIC: electrons from a polarized gun are accelerated in a relatively small 3-pass, 4 GeV, energy recovery linac (ERL) to collide with 100 A GeV protons or ions at a single RHIC IP.
- eRHIC: the original ERL is modified into the pre-injector for a much larger ERL, consisting of a stack 4 to 6 vertically separated rings extending around the full RHIC tunnel. The electrons make several accelerating and decelerating passes through two superconducting linacs at multiple energy stages up to 30 GeV. Among several innovations in the design is the use of the coherent electron cooling for the hadron beams.

4.2. MEIC and ELIC

In contrast to BNL, JLAB already has a recirculating electron accelerator, the 12 GeV CEBAF, and proposes to add hadron beams, resulting in a very different facility from eRHIC. Protons and light ions would be supplied by a new hadron accelerator complex, consisting of ion sources, a superconducting linac, and pre-booster. In the initial stage, dubbed MEIC, the two types of beam are collided at 4 vertically-crossing IPs close to the centre of a pair of stacked figure-eight rings. Energies of the e-p collisions could be 3-11 GeV on 12-60 GeV depending on whether the hadron ring is chosen to be superconducting. In the later ELIC stage, a larger hadron ring could provide energies up to 250 GeV.

By applying several innovative concepts (high bunch collision frequency, small charge and spot size, crab crossing, …) MEIC already aims for very high luminosity ($4 \times 10^{34}$ cm$^{-2}$s$^{-1}$) at 3 on 60 GeV and the upgrade path to the full ELIC luminosity is even more ambitious. The goal of the JLab team is to develop a detailed and credible machine design, followed by a long term R&D program to reach ultra high luminosity with high polarization levels of light ions and electrons.

LHeC

Although CERN previously had the e$^\pm$ beams of LEP in what is now the LHC tunnel, the magnets were removed to make way for those of the LHC. Moreover, the lepton injectors were either re-purposed or modified to avoid compromising their performance for high intensity proton beams. The LHeC proposal to collide polarized e$^\pm$ beams with the hadron beams (p, Pb, D, …) of the LHC therefore requires new lepton sources and injectors. At present, two concepts for the LHeC meeting the overall limit of 100 MW power consumption are under consideration:

- Ring-ring: a new ring of lightweight low-field magnets mounted above the LHC in the same tunnel with a few km of bypasses around the LHC experiments, a substantial superconducting RF system to take high current beams up to $E_e \approx 60$ GeV. Given the experience with LEP, where polarized levels of ~50% were demonstrated, this can be regarded as a fairly conservative design; the main challenges are related to finding space for the ring and its bypasses in the same tunnel as the LHC. This has, for example, imposed an unconventional design of the basic FODO arc cell.
- Linac-ring: either a pulsed linac or an ERL, in various versions, providing lepton energies in the range of 60-140 GeV.

Either design would provide a spectacular increase in accessible energies and low Bjorken *x* and is intended to collide e-p ($L \approx 10^{33}$ cm$^{-2}$s$^{-1}$) without disturbing the parallel high-luminosity p-p collisions in the LHC. As an e-A collider ($L_{rN} \approx 10^{32}$ cm$^{-2}$s$^{-1}$) it could work stand-alone or in parallel with a continued A-A programme in the LHC. The goal of the multi-institute LHeC collaboration is to produce a conceptual design report in 2011 and choose between these two concepts soon afterwards.

5. Outlook

The remarkable progress of high energy nuclear physics in the past decade (and before) has been powered by particle accelerator technology, developed in the first instance for elementary particle physics, but extended and enriched by the more elaborate beam requirements of nuclear physics.

A decade of extraordinary achievement and innovation in accelerator physics at RHIC has led to major results in nuclear physics. Furthermore the pace of innovation has not slowed down: we can expect continuing performance improvements of this remarkably flexible collider for many years to come.

The next decade will see three remarkably powerful detector systems brought to bear upon the exploration of a vastly extended energy frontier at the LHC. Discussion of a future beyond about 2020 is premature until the first physics results are in. However my personal view is that it would be surprising if we are not motivated to upgrade the heavy-ion luminosity (via cooling systems, perhaps) as is already happening for the p-p programme.

In parallel with the high energy collider programmes, continued extension of the capabilities of RHIC, the SPS, FAIR and other facilities will explore other regions of the QCD phase diagram.

Current proposals for e-p/A colliders exploit a diversity of accelerator design concepts and aim to provide similarly diverse physics capabilities. Both aspects, feasibility of these very complex facilities and desirability, must be considered

## Acknowledgements

This talk sketched some aspects of the work of many people, over many years, in several institutes around the world. Particular thanks for help and use of material in preparing this talk and paper are due to G. Bellodi, I. Ben-Zvi, R. Bruce, H. Burkhardt, C. Carli, M. Ferro-Luzzi, W. Fischer, M. Fitterer, M. Gazdzicki, P. Giubellino, B. Holzer, K. De Jager, M. Klein, D. Kuchler, M. Lamont, V. Litvinenko, D. Manglunki, S. Maury, A. Morsch, T. Roser, C. Salgado, J. Schukraft, P. Steinberg, D. Tommasini, U.A. Wiedemann, B. Wosiek, B. Wyslouch and Y. Zhang.

## References

All the following accelerator conference references (eg, [1,10]) can be found via http://www.jacow.org/.

[1] P. Asboe-Hansen et al, *Acceleration and stacking of deuterons in the CERN PS and ISR*, IEEE Trans. Nucl. Sci., Vol. NS-24, No. 3 (1977) 1557.
[2] M. Boutheon et al, *Acceleration and stacking of α-particles in the CERN Linac, PS and ISR,* IEEE Trans. Nucl. Sci., Vol. NS-28, No. 3 (1981) 2049.
[3] A. W. Chao, M. Tigner, *Handbook of Accelerator Physics and Engineering*, World Scientific, Singapore, 2006.
[4] J.M. Jowett et al, *Limits to the performance of the LHC with ion beams,* Proc. EPAC 2004, Lucerne 2004.
[5] R.Bruce et al, Time evolution of the luminosity of colliding heavy-ion beams in BNL Relativistic Heavy Ion Collider and CERN Large Hadron Collider, Phys. Rev. ST Accel. Beams **13**, 091001 (2010).
[6] A. J. Baltz, M. J. Rhoades-Brown, and J. Weneser, *Heavy-ion partial beam lifetimes due to Coulomb induced processes,* Phys. Rev. E 54, 4233 (1996).
[7] A. Morsch, I. Pshenichnov, *LHC Experimental Conditions,* ALICE Inernal Note 2002-034, 2002.
[8] C. Montag et al, *RHIC Performance as a 100 GeV polarized proton collider in Run-9*, Proc. IPAC'10, Kyoto, 2010.
[9] W. Fischer, *RHIC luminosity upgrade program,* Proc.IPAC'10, Kyoto, 2010.
[10] K.A. Brown et al, RHIC *Performance for FY10 200 GeV Au+Au Heavy Ion Run*, Proc. IPAC'10, Kyoto, 2010.
[11] M. Blaskiewicz et al, *Operational Stochastic Cooling in the Relativistic Heavy-Ion Collider*, Phys. Rev. Lett. 100, 174802 (2008).
[12] *LHC Design Report Vol I: The LHC Main Ring* Chap. 21 and *Vol III The LHC Injector Chain*, Chaps. 32-38, CERN-2004-003.
[13] J. M. Jowett, *The LHC as a nucleus–nucleus collider*, J. Phys. G: Nucl. Part. Phys. 35 104028 (2008)
[14] D. Manglunki et al, Ions for LHC: towards completion of the injector chain, Proc.EPAC'08, Genoa, 2008.
[15] S. Myers, *LHC commissioning and first operation,* Proc.IPAC'10, Kyoto, 2010.
[16] S.R. Klein, *Localized beampipe heating due to e- capture and nuclear excitation in heavy ion colliders*, Nucl. Instrum. Methods Phys. Res., Sect. A, **459** (2001) 51.
[17] R. Bruce et al, *Beam losses from ultraperipheral nuclear collisions between $^{208}Pb^{82+}$ ions in the Large Hadron Collider and their alleviation,* Phys. Rev. ST Accel. Beams **12**, 071002 (2009)
[18] T. Satogata et al, *Commissioning of RHIC deuteron-gold collisions*, Proc. PAC 2003, Portland, 2003.
[19] J.M. Jowett and C. Carli, *The LHC as a proton-nucleus collider,* Proc. EPAC2006, Edinburgh, 2006/
[20] V. Litvinenko, *Future Electron-Hadron Colliders*, Proc.IPAC'10, Kyoto, 2010.
10